\documentclass[journal]{IEEEtran}
\usepackage[utf8]{inputenc}
\usepackage{amsmath,amssymb,amsfonts}
\usepackage{algorithm} 
\usepackage{epstopdf}
\usepackage{subcaption} % 使用 subcaption 来处理子图
\usepackage{caption}    % 使用 caption 宏包控制 caption 样式
\usepackage{textcomp}
\usepackage{bm}
\usepackage{amsmath,amsfonts}
\usepackage{algorithmic}
\usepackage{array}
\usepackage{textcomp}
\usepackage{stfloats}
\usepackage{url}
\usepackage{verbatim}
\usepackage{graphicx}
\usepackage{booktabs}
\usepackage{balance}
\usepackage{multirow}
\usepackage{diagbox}
\usepackage{float}
\usepackage{xcolor}
\def\BibTeX{{\rm B\kern-.05em{\sc i\kern-.025em b}\kern-.08em
		T\kern-.1667em\lower.7ex\hbox{E}\kern-.125emX}}
\usepackage{balance}
\vspace{-10.pt}
\begin{document}
	\title{\huge Rotatable RIS-Assisted Edge Computing: Orientation, Task Offloading, and Resource Optimization}
	\author{Bin Li, Dongdong Yang, and Lei Liu
	
	\thanks{Bin Li and Dongdong Yang are with the School of Computer Science, Nanjing University of Information Science and Technology, Nanjing 210044, China (e-mail: bin.li@nuist.edu.cn; 202312200024@nuist.edu.cn).}
\thanks{Lei Liu is with the Guangzhou Institute of Technology, Xidian University, Guangzhou 510555, China (e-mail: tianjiaoliulei@163.com).}	
	}
	
	\setlength{\parskip}{0pt} % 段落间距
	\maketitle
	
\vspace{-15.pt}
	\begin{abstract}
	The rotatable reconfigurable intelligent surface (RIS) can enhance mobile edge computing (MEC) performance by optimizing its orientation to improve the gain of received and transmitted signals. This correspondence investigates a rotatable RIS-assisted MEC system, aimed at minimizing energy consumption for multiple moving user equipment (UEs) through the joint design of RIS orientation, discrete phase shift, computation resource allocation, transmitting power and task offloading strategies. Considering the mobility of UEs, this problem is formulated as a sequential decision-making across multiple time slots. To address this challenge, a soft actor-critic (SAC)-based algorithm is proposed to optimize RIS orientation, phase shift and task offloading strategies, while computation resource allocation and transmitting power are determined based on the actions. Numerical results demonstrate that the proposed scheme exhibits superior convergence and performance compared to benchmarks. Additionally, the rotatable RIS scheme reduces total energy consumption by up to 47.3\% compared to the fixed RIS, enhancing MEC system performance.
	\end{abstract}
	\begin{IEEEkeywords}
	Mobile edge computing, rotatable reconfigurable intelligent surface, orientation, soft actor-critic, user mobility.
	\end{IEEEkeywords}
	
\vspace{-10.pt}	
	\section{Introduction}
	Reconfigurable intelligent surface (RIS) has gained recognition as a key technology in 6G for its ability to improve channel quality, along with its passivity, reconfigurability and cost-effectiveness \cite{9847080}. By integrating RIS with mobile edge computing (MEC), the advantages of RIS can be fully leveraged to significantly enhance offloading links, providing substantial benefits in MEC systems. For example, \cite{10128149} introduced an optimization strategy in RIS-assisted MEC non-orthogonal multiple access networks, using $\epsilon$-constraint approach and a successive convex approximation-based algorithm to minimize the offloading delay and maximize transmission rate. To further enhance energy and computation efficiency, \cite{hu2021reconfigurable} proposed a novel algorithm based on deep learning. In addition, considering the task offloading security, \cite{xu2023deep} explored the benefits of RIS in MEC-enabled networks, utilizing a deep deterministic policy gradient-based algorithm to maximize the weighted sum secrecy computation efficiency among devices.
	
	Given that the antenna's receiving and transmitting gain is affected by the angle of wave incidence and transmission \cite{cheng2022ris}, the RIS orientation optimization becomes a crucial aspect worth investigating. Nevertheless, existing studies have predominantly focused on performance analysis and optimization in fixed or single-UE scenarios, which may not adequately address the complexities of scenarios involving multiple moving UEs in MEC systems. For instance, \cite{9903366} evaluated the impact of RIS position and orientation in a single mobile UE environment, highlighting that the orientation significantly influences the received power. Similarly, \cite{10008708} demonstrated that optimizing RIS orientation can achieve higher average throughput in low earth orbit satellite communication systems.
	
	To the best of our knowledge, the potential of rotatable RIS in MEC systems has not been thoroughly investigated in existing studies. Notably, for rotatable RIS-assisted MEC systems, the phase shift is highly coupled with orientation and offloading strategies, and typically discrete under practical constraints, making it challenging for traditional algorithms to identify optimal solutions \cite{wang2024power}. Nevertheless, deep reinforcement learning (DRL) offers innovative approaches to addressing such optimizing problems in time-varying scenarios. For example, \cite{zhu2022drl} proposed a soft actor-critic (SAC)-based joint beamforming and base station (BS)-RIS-UE association design to maximize the sum-rate in millimeter wave communication systems. Similarly, in the RIS-assisted MEC context, \cite{xu2024deep} developed a twin delayed deep deterministic policy gradient-based algorithm for computation rate maximization.

	Motivated by the DRL methods, a policy-based SAC algorithm is adopted to tackle the problem of minimizing energy consumption in the rotatable RIS-assisted MEC system with multiple moving UEs, where RIS orientation, discrete phase shift, computation resource allocation, transmitting power, and task offloading strategies are jointly involved. In particular, the SAC algorithm employs a stochastic strategy with entropy regulation as its main feature, enabling it to learn near-optimal policies with minimal environmental information.
	
	\section{System Model and Problem Formulation}\label{s:sys}
\begin{figure}[t]
		\centering
		\includegraphics[width=0.45\textwidth]{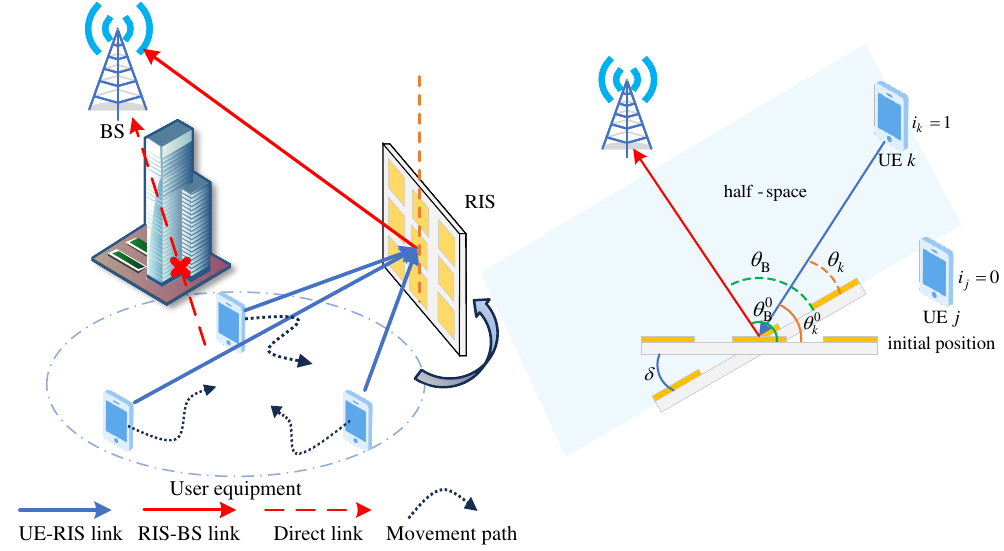}
		\caption{System model and the corresponding angle variations during RIS rotation.}
		\label{fig:sys-model}
	\end{figure}

	As illustrated in Fig. \ref{fig:sys-model}, we consider a rotatable RIS-assisted MEC system, consisting of $K$ single-antenna moving UEs, a rotatable RIS with $N$ RIS elements, and a single-antenna BS integrated with an MEC server. The set of UEs and RIS elements are denoted by $\mathcal{K}=\{ 1,...,k,...,K\}$ and $\mathcal{N}=\{ 1,...,n,...,N\}$, respectively. Since the direct links between the BS and UEs are prone to be blocked, the RIS is strategically deployed to maintain line-of-sight (LoS) paths between them. Additionally, the RIS can dynamically adjust its orientation according to the relative positions of the BS and UEs. 
	
\vspace{-10.pt}	
	\subsection{Communication Model}
	In this correspondence, we assume that all UEs operate within a task cycle of duration $T$, equally divided into $Q$ intervals, each with length $\tau = T/Q$, where the set of time slots is represented as $\mathcal{Q}=\{ 1,\ldots,q,\ldots,Q\}$.  During each time slot $q$, the channels from UE $k$ to the RIS and from the RIS to BS are denoted by ${\bm{h}_{k,{\rm{R}}}}[q]$ and $\bm{v}_{\rm{R},\rm{B}}[q]$, respectively. These overall channels are determined by the individual channels between UE $k$ and each RIS element $h_{k,\rm{R}}^n[q]$, as well as the channels between each RIS element and the BS $v_{\rm{R},\rm{B}}^n[q]$, both modeled as Rician distribution \cite{cheng2022ris}:
	\begin{equation}
	\scalebox{0.81}{$
		\begin{aligned}
			h_{k,\rm{R}}^n[q] &= \sqrt{\frac{\rho_0}{{d_{k,\rm{R}}^n[q]}^{\alpha_1}}} 
			\left( 
			\sqrt{\frac{K_1}{1 + K_1}} \bar{h}_{k,\rm{R}}^n[q] + 
			\sqrt{\frac{1}{1 + K_1}} \tilde{h}_{k,\rm{R}}^n[q] 
			\right),
		\end{aligned}
		$}
	\end{equation}
	\begin{equation}
	\scalebox{0.81}{$
		\begin{aligned}
			v_{\rm{R},\rm{B}}^n[q] &= \sqrt{\frac{\rho_0}{{d_{\rm{R},\rm{B}}^n[q]}^{\alpha_2}}} 
			\left( 
			\sqrt{\frac{K_2}{1 + K_2}} \bar{v}_{\rm{R},\rm{B}}^n[q] + 
			\sqrt{\frac{1}{1 + K_2}} \tilde{v}_{\rm{R},\rm{B}}^n[q] 
			\right),
		\end{aligned}
		$}
	\end{equation}
	where ${\rho_0}$ denotes the path-loss factor at a reference distance of 1 meter; $d_{k,\rm{R}}^n[q]$ ($d_{\rm{R},\rm{B}}^n[q]$) signifies the distance between UE $k$ (BS) and the $n^{\rm th}$ RIS element; $\alpha_1$ and $\alpha_2$ represent the path loss exponents; $K_1$ and $K_2$ are the Rician factors. The LoS components are given by ${{\bar h}_{k,\rm{R}}^n}[q]=\exp \left(-j\frac{2\pi}{\lambda} d_{k,\rm{R}}^n[q]\right)$ and ${{\bar v}_{\rm{R},\rm{B}}^n}[q]=\exp \left(-j\frac{2\pi}{\lambda} d_{\rm{R},\rm{B}}^n[q]\right)$, while the non-LoS components ${{\tilde h}_{k,\rm{R}}^n}[q]$ and ${{\tilde v}_{\rm{R},\rm{B}}^n}[q]$ follow a circularly symmetric complex Gaussian distribution with zero mean and unit variance. Since the LoS paths between UEs and the BS are blocked, the channel from UE $k$ to the BS $h_{k,\rm{B}}$ is modeled as Rayleigh fading channel. %${h_{k,{\rm{B}}}} = \sqrt {{\rho _0}/d_{k,{\rm{B}}}^{{\alpha _3}}} {\tilde h_{k,{\rm{B}}}}[q]$, with $\alpha_3$ denoting the path loss exponent, and $d_{k,\rm B}$ representing the distance between UE $k$ and the BS.
	
	The angle between UE $k$ and the RIS plane, as well as between the RIS plane and BS are denoted by ${\theta _k[q]}$ and ${\theta_{\rm B}[q]}$, respectively, which can be calculated as follows
	 \begin{equation}
		{\theta _{k(\rm{B})}[q]} =  {\theta _{k(\rm{B})}^0[q] - \delta[q] },
	\end{equation}
	where $\delta [q]$ denotes the rotation angle, and $\theta_{k(\rm{B})}^0[q]=\left\langle n_0, n_{k(\rm{B})}[q] \right\rangle$ represents the angle between UE $k$ (BS) and the initial orientation RIS, with $n_{k(\rm{B})}[q]$ indicating the direction between UE $k$ (BS) and the RIS, and $n_0$ denoting the initial RIS orientation. To ensure successful task offloading to the BS, $\theta_{\rm{B}}[q]$ must be maintained within the range $\theta_{\rm{B}}[q] \in [0,\pi]$. Thus, the constraint on the rotation angle $\delta[q]$ is given by
	\begin{equation}
		\delta[q] \in [\theta_{\rm{B}}^0-\pi,\theta_{\rm{B}}^0].
	\end{equation}
	
	Considering the rotation of the RIS, some UEs may fall outside the reflection half-space as it rotates. Therefore, $i_k$ is introduced to indicate whether UE $k$ remains within the half-space, defined by
	\begin{equation}
		{i_k}[q] = \left\{ {\begin{array}{*{20}{c}}
				{1,}&{{\theta _k}[q] \in [0,\pi ]},\\
				{0,}&{{\rm{otherwise.}}\quad\;}
		\end{array}} \right.
	\end{equation}
	
	The location of UE $k$ at time slot $q$ is defined as ${l_k}[q] = [{x_k}[q],{y_k}[q],0]$, and updated following the Guass-Markov random model \cite{bin2024sac}. 
	The reflection coefficient matrix of the RIS in time slot $q$ is represented by the diagonal matrix $\mathbf{\Phi} [q] = {\rm{diag}}({e^{j{\phi _1}[q]}},\ldots,{e^{j{\phi _n}[q]}},\ldots,{e^{j{\phi _N}[q]}})$, where ${e^{j{\phi _n}[q]}}$ denotes the phase shift of the $n^{\rm{th}}$ RIS element in time slot $q$. It is worth noting that discrete phase shift must be considered due to practical hardware limitations. Therefore, the $b$-bit discrete phase shift set is characterized as
	\begin{equation}
		\mathcal{F} = \{ {e^{j{\phi _n}[q]}}|{\phi _n}[q] \in \{ 0,{2^{1 - b}}\pi ,...,(2 - {2^{1 - b}})\pi \} \}.
	\end{equation}
	
	Since the BS and UEs are situated in the far-field region of the RIS, we assume that the angle between UE $k$ (BS) and each RIS element is equal. Thus, the gain of the RIS for UE $k$ is equivalent to
	\begin{equation}
		\scalebox{0.93}{$
			\boldsymbol{\xi}_k[q] = G_k[q] G_{\rm{B}}[q] \mathbf{\Phi}[q] \triangleq D_{\rm m}^2 F\left(\Theta_k[q]\right) F\left(\Theta_{\rm{B}}[q]\right) \mathbf{\Phi}[q],
		$}
	\end{equation}
	where $D_{\rm m}$ is the maximum directivity of the RIS, $G_{k}[q]$ denotes the receiving gain from UE $k$ to the RIS, $G_{\rm{B}}[q]$ represents the transmitting gain from the RIS to BS, and $F\left(\Theta\right)$ signifies the normalized power radiation pattern of the RIS \cite{cheng2022ris},  modeled using an exponential-Lambertian radiation pattern parameterized by $z$, which is given by
	\begin{equation}
		F\left( \Theta \right) = \cos^z\left( \upsilon \right) \cos^z\left( \rho \right),  \upsilon \in \left[ 0, 2\pi \right], \, \rho \in \left[ 0, \pi \right],
	\end{equation}
	where $\upsilon$ and $\rho$ denote the azimuth and elevation angles between UE $k$ (BS) and RIS, respectively. The elevation angle $\rho$ can be ignored since all devices are located at the same altitude. According to equations (5), (7) and (8), the gain of the RIS can be rewritten as
	\begin{equation}
		\boldsymbol{\xi}_k[q] = i_k[q]D_{\rm{m}}^2{\sin^z}(\theta_k[q]){\sin^z}(\theta_{\rm{B}}[q])\mathbf{\Phi}[q].
	\end{equation}
	
	Furthermore, we define ${h_k}[q] = {\boldsymbol{v}_{\rm{R},\rm{B}}}{[q]^H}\boldsymbol{\xi}_k[q] {\boldsymbol{h}_{k,\rm{R}}}[q] + {h_{k,\rm{B}}}[q]$. By utilizing the orthogonal frequency division multiple access protocol for task offloading, the achievable rate of UE $k$ in time slot $q$ is expressed as
	\begin{equation}
		{R_k}[q] = {B_k}{\log _2}\left( {1 + {p_k}[q]{{\left| {{h_k}[q]} \right|}^2}/{\sigma ^2}} \right),
	\end{equation}
	where ${B_k} = B/K$ denotes the bandwidth allocated to UE $k$, $B$ is the total bandwidth, $\sigma ^2$ signifies the Gaussian noise power, and ${p_k}[q]$ represents the transmitting power of UE $k$.
	
	\subsection{Computation Model}
	Each UE $k$ has a task ${M_k} = \{ {D_k},{C_k}\}$ to handle at the beginning of cycle $T$, where ${D_k}$ is the task size and ${C_k}$ denotes the required CPU cycles per bit. UEs perform partial offloading, splitting tasks between local and edge computing. Considering that the BS is equipped with a high-performance server, its energy consumption can be neglected. Therefore, the system's energy consumption is primarily concentrated in task offloading and local computing.
	
	 \subsubsection{Task offloading}Since the process of UEs offloading tasks to the BS and the MEC server executing the offloaded tasks requires time, we assume that the MEC server does not execute tasks in the first time slot, and UEs do not offload tasks in the $Q$-th time slot. UE $k$ offloads ${\alpha _k}[q] \in [0,1]$ of task size ${D_k}$ in time slot $q$, satisfying $\eta_k=\textstyle \sum\nolimits_{q = 1}^Q {{\alpha _k}[q]}  \le 1$, so that $(1 - \eta_k){D_k}$ will be processed locally. The time and energy consumption for task offloading are described as
	\begin{equation}
		t_k^{\rm{off}}[q] = {\alpha _k}[q]{D_k}{({R_k}[q])^{ - 1}},\forall q \in\mathcal Q,
	\end{equation}
	\begin{equation}
		E_k^{{\rm{off}}}[q] = t_k^{{\rm{off}}}[q]{p_k}[q],\forall q \in \mathcal Q.
	\end{equation}
	After receiving the offloaded tasks, the MEC server executes them in parallel. Let $f_k^{\rm{e}}$ denote the computation resource allocated to UE $k$ during cycle $T$, subject to $\sum\nolimits_{k = 1}^K {f_k^{\rm{e}}}  \le f_{{\rm{total}}}^{\rm{e}}$, where $f_{{\rm{total}}}^{\rm{e}}$ represents the total computation resource of the MEC server. Thus, the offloaded computation time of UE $k$ is calculated by
	\begin{equation}
		t_k^{\rm{e}} = {D_k}{C_k}{\eta_k}/f_k^{\rm{e}}.
	\end{equation}
	\subsubsection{Local processing}The local computation time and energy consumption of UE $k$ during cycle $T$ are given by
	\begin{equation}
		t_k^{{\rm{loc}}} = {D_k}{C_k}(1 - {\eta_k})/f_k^{{\rm{loc}}},
	\end{equation}	
	\begin{equation}
		E_k^{{\rm{loc}}} = {c_k^{{\rm{loc}}}}{\left( {f_k^{{\rm{loc}}}} \right)^2}{D_k}{C_k}(1 - {\eta_k}),
	\end{equation}
	where $f_k^{{\rm{loc}}}$ and $c_k^{{\rm{loc}}}$ represent the CPU frequency and effective capacitance coefficient of UE $k$, respectively.
	\subsection{Problem Formulation}
	Our objective is to minimize the total energy consumption by jointly optimizing RIS orientation, discrete phase shift, computation resource allocation, transmitting power, and offloading strategies, while adhering to constraints on latency and computation resource. Accordingly, the energy consumption minimization problem in the rotatable RIS-assisted MEC system is formulated as
	\begin{subequations}
		\begin{flalign}
			&\mathop {\min }\limits_{\delta ,{\phi _n},{\alpha _k},{p_k},f_k^{{\rm{loc}}},f_k^{\rm{e}}} \;\sum\limits_{k = 1}^K {{E_k}} &&\\
			&\quad{\rm{s}}{\rm{.t}}{\rm{.}}\, \delta[q] \in [\theta_{\rm{B}}^0-\pi,\theta_{\rm{B}}^0], \forall q \in {\cal Q}, &&\\
			&\quad\quad\text{ }  {\phi _n}[q] \in {\cal F},\forall n \in {\cal N},\forall q \in {\cal Q}, &&\\
			&\quad\quad\text{ } f_k^{{\rm{loc}}} \in [0,f_{\max }^{{\rm{loc}}}],\forall k \in \mathcal K, &&\\
			&\quad\quad\text{ } \sum\limits_{k = 1}^K {f_k^{\rm{e}}}  \le f_{{\rm{total}}}^{\rm{e}},\forall k \in \mathcal K, &&\\
			&\quad\quad\;\;{p_k}[q] \in [0,{p_{\max }}],\forall k \in \mathcal K,\forall q \in \mathcal Q, &&\\
			&\quad\quad\text{ } \max \{ t_k^{{\rm{loc}}},t_k^{{\rm{off}}},t_k^{\rm{e}}\}  \le T,\forall k \in \mathcal K, &&\\
			&\quad\quad\;\; \sum\limits_{q = 1}^Q {{\alpha _k}[q]}  \le 1,{\alpha _k}[q] \in [0,1],\forall k \in \mathcal K,\forall q \in \mathcal Q, &&
			%&\quad\quad\;\, \sum\limits_{q = 1}^Q {{\alpha _k}[q]}  \le 1,\forall k \in \mathcal K, &&
			%&\quad\quad\text{ } \left| {\delta [q] - \delta [q-1]} \right| \le \frac{\pi }{4},\forall q \in \mathcal Q \\
			%&\quad\;\; (4),(6),&&
		\end{flalign}
	\end{subequations}
	where ${E_k} = E_k^{{\rm{loc}}} + \sum\nolimits_{q = 1}^{Q - 1} {E_k^{{\rm{off}}}[q]} $ and $t_k^{{\rm{off}}} = \sum\nolimits_{q = 1}^{Q - 1} {t_k^{{\rm{off}}}[q]}$. Constraint (16b) limits the range for RIS orientation, ensuring the BS remains within the reflective half-space of the RIS. Constraint (16c) restricts the RIS phase shift to the predefined discrete set $\cal F$. Constraint (16d) imposes a maximum local computation resource limitation, preventing the local computation load at each UE from surpassing its capacity. Constraint (16e) guarantees that the total allocated computation resource of MEC server does not exceed the total resource. Constraint (16f) defines the range of the transmission power. Constraint (16g) ensures that all tasks can be completed within cycle $T$. Constraint (16h) prevents the offloaded tasks from exceed the total task size.

	The major challenges in solving problem (16) arise from the strong coupling of variables, such as RIS orientation and phase shift, and the non-convexity of the objective function with respect to the involved variables. Additionally, UEs mobility introduces uncertainties in the optimization process. 
	
	\section{Problem Solution}\label{pro:s}
	In this section, we design RIS orientation, phase shift, and task offloading strategies based on SAC algorithm. Subsequently, the transmitting power and computation resource allocation are determined in the DRL environment according to the actions.
	
	\subsection{MDP Formulation}
	In order to implement DRL, we first define MDP which serves as a fundamental framework to model and solve sequential decision-making problems in a stochastic environment. The components of MDP include state space $\mathcal{S}$, action space $\mathcal{A}$, state transition probability function $\mathcal{P}$, and reward $\mathcal{R}$.
	
	\subsubsection{State}
	The environment in time slot $q$ consists of three parts: the distances between UEs and RIS $\boldsymbol{d}[q] = \{ {d_{k,\rm R}},\forall k \in \mathcal K\} $, the angles between UEs and RIS $\boldsymbol {\theta}[q]  = \{ {\theta _k}[q],k \in \mathcal K\} $, and the cumulative ratios of offloaded tasks $\boldsymbol {\alpha}^{\rm {cu}} [q] = \left\{ {\sum\nolimits_{i = 1}^{q - 1} {{\alpha _k}[i]} ,k \in \mathcal K} \right\}$. Therefore, the state space is given by
	\begin{equation}
		{s_q} = \left\{ {\boldsymbol d[q],\boldsymbol\theta [q],\boldsymbol\alpha^{\rm{cu}} [q],\forall q \in \mathcal Q} \right\}.
	\end{equation}
	
	\subsubsection{Action}
	The action space of the formulated MDP includes the rotation angle $\delta[q]$ and phase shift of the RIS, as well as UEs' task offloading strategies $\boldsymbol\alpha [q] = \left\{ {{\alpha _k}[q],k \in \mathcal K} \right\}$. Considering that the phase shift is discrete, the phase shift space $[0,2\pi )$ is divided into a $b$-bit discrete phase shift set $\boldsymbol\phi [q] = \left\{ {{\phi _n}[q] \in [0,2 - {2^{1 - b}}],\forall n \in \mathcal N} \right\}$. Therefore, the action space is defined as
	\begin{equation}
		{a_q} = \left\{ {\delta [q],\boldsymbol\phi [q],{\boldsymbol\alpha}[q],\forall q \in \mathcal Q} \right\}.
	\end{equation}
	
	\subsubsection{Reward}
	Our objective is to minimize the total energy consumption, while the goal of the agent is to maximize the reward. Consequently, the reward should be inversely related to the objective function. Additionally, to enforce constraints (16b)-(16i), penalties are introduced if any of these constraints are violated. Thus, the reward is given by
	\begin{equation}
		\arraycolsep=1.4pt % 调整列间距
		r(s_q, a_q) = \left\{
		\begin{array}{ll}
			- \sum\limits_{k = 1}^K E_k^{\text{off}}[q] - P_1, & q \leq Q - 1, \\
			- \sum\limits_{k = 1}^K (E_k^{\text{loc}} + E_k^{\text{off}}) - P_2, & q = Q - 1,
		\end{array}
		\right.
	\end{equation}
	where ${P_1} = {K^{{\rm{un}}}}W + {P_{{\rm{theta}}}}$ denotes the penalty for unreasonable offloading ratios, suboptimal orientation and discrete phase shift. Here,  $K^{{\rm{un}}}$ denotes the number of unsatisfactory offloading ratios, and $W$ is a positive constant, $P_{\rm{theta}}=W$ represents the penalty applied when the performance of random phase shift and orientation surpasses these determined by the policy network. %$P_2$ denotes the penalty for excessive rotation angle, which will equal to $W$ when violate the constraint (18h). 
	For the $(Q-1)$-th time slot, the penalty is set as ${P_2} = \left( {\sum\nolimits_{k = 1}^K {f_k^{\rm{e}}}  - f_{{\rm{total}}}^{\rm{e}}} \right)W + {P_1}$, where $\left( {\sum\nolimits_{k = 1}^K {f_k^{\rm{e}}}  - f_{{\rm{total}}}^{\rm{e}}} \right)W$ accounts for the penalty due to inappropriate edge computation resource allocation.

	\subsection{Optimization for Computation Resource Allocation and Transmitting Power}
	In this section, we describe the method for determining the computation resource allocation and transmitting power based on the actions taken at each training step.
	\subsubsection{Computation resource allocation}
	According to equation (15), the local computation energy $E_k^{\rm {loc}}$ increases with $f_k^{\rm {loc}}$. Therefore, when $f_k^{\rm {loc}}$ reaches its minimum value, $E_k^{\rm {loc}}$ is minimized. From equation (14), when local computation time $t_k^{\rm {loc}}=T$, $f_k^{\rm {loc}}$ achieves its optimal value, which can be expressed as
	\begin{equation}
		{ {f_k^{{\rm{loc}}}} ^*} = {D_k}{C_k}\left( {1 - {\eta _k}} \right)/T.
	\end{equation}
	
	Similarly, to execute the offloaded tasks within $Q-1$ time slots, the optimal computation resource allocation is given by
	\begin{equation}
		{ {f_k^{\rm{e}}}^*} = {D_k}{C_k}{\eta _k}/\left( {\left( {Q - 1} \right)\tau } \right).
	\end{equation}
	
	\subsubsection{Transmitting power}
	Based on equation (11), we can demonstrate that $t_k^{\rm {off}}[q]$ is inversely related to $p_k[q]$. Given that the offloading time $t_k^{{\rm{off}}}[q] \in [0,\tau]$, the following constraint must be satisfied:
	\begin{equation}
		p_k[q] \ge {\sigma ^2}\left( {{2^{{\alpha _k}[q]{D_k}{{\left( {\tau {B_k}} \right)}^{ - 1}}}} - 1} \right){\left| {{h_k}[q]} \right|^{ - 2}} = \hat p_k[q],
	\end{equation}
	where $t_k^{\rm {off}}[q] \le \tau$ and $\hat p_k[q] \le p_{\rm{max}}$ are enforced by the penalty $P_1$.  Combining equations (11), (12) and (22), the problem of minimizing transmitting energy consumption through optimizing transmitting power is formulated as
	\begin{subequations}
			\begin{align}
				&\mathop {\min }\limits_{{p_k[q]}} \;E_k^{{\rm{off}}}[q] = \frac{{{\alpha _k}[q]{D_k}[q]{p_k}[q]}}{{B_{k}{{\log }_2}\left( {1 + {p_k}[q]{{\left| {{h_k}[q]} \right|}^2}/{\sigma ^2}} \right)}}\\
				&{\rm{s}}{\rm{.t}}{\rm{.}}\;{\hat p_k[q] \le p_k[q] \le {p_{\max }}}.
			\end{align}
	\end{subequations}
	
	To minimize this fractional problem, Dinklebach's transform is applied to decouple and solve it, reformulating the objective (23a) as
	\begin{equation}
		\scalebox{0.92}{$
			\mathop {\min }\limits_{{p_k}[q]} \;{\alpha _k}[q]{D_k}[q]{p_k}[q] - {y_k}{B_k}{\log _2}\left( {1 + {p_k}[q]{{\left| {{h_k}[q]} \right|}^2}/{\sigma ^2}} \right),
		$}
	\end{equation}
	with a new auxiliary variable $y_k$, iteratively updated by
	\begin{equation}
		{{y_k}^{t + 1}}{\rm{ = }}\frac{{{\alpha _k}[q]{D_k}[q]{p_k}{{[q]}^t}}}{{B_{k}{{\log }_2}\left( {1 + {p_k}{{[q]}^t}{{\left| {{h_k}[q]} \right|}^2}/{\sigma ^2}} \right)}},
	\end{equation}
	where $t$ is the iteration index. Convergence is ensured by alternatively updating $y$ according to (25) and solving for $p_k[q]$ in (24), as $y$ is non-increasing with each iteration \cite{shen2018fractional}. The optimal transmitting power ${{{p_k}[q]}^*}$ can be obtained after successive iterations.
	
	\subsection{SAC-Based Algorithm}
	The much awaited application of DRL frameworks has progressed slowly in practice, primarily due to the relatively poor sampling efficiency and brittle convergence \cite{zhao2022simultaneously}. To address these challenges, SAC was proposed, leveraging the maximum entropy framework to achieve higher sampling-efficient training. 
	
	The objective of conventional DRL frameworks is to maximize the long-term return starting from the initial state. However, in SAC, an entropy term is incorporated into the objective function to encourage exploration, given by
	\begin{equation}
		\sum\limits_q \mathbb{E}_{(s_q, a_q) \sim \rho_\pi} \left[ \gamma^{q - 1} r(s_q, a_q) + \alpha \mathcal{H}(\pi(\cdot| s_q)) \right],
	\end{equation}
	where $\mathcal H(\pi(\cdot | s_q)) = - E_{a \sim \pi(\cdot | s_q)} \log_2 \pi(a | s_q)$ represents the entropy of the policy distribution, and the temperature parameter $\alpha$ controls the weight of the entropy, indicating how stochastic the optimal policy ${{\rm{\pi }}^*}$ is.% Given the dynamic nature of the reward, $\alpha$ is necessary to be adjusted automatically. According to \cite{bin2024sac}, the optimal dual variable $\alpha_q^{*}$ at each step is expressed as
	% \begin{equation}
	% 	\scalebox{0.85}{$
	% 	\alpha_q^* = \mathop{\arg\min}_{\alpha_q} \mathbb{E}_{a_q \sim \pi_q^*} \left[ -\alpha_q \log_2 (\pi_q^* (a_q \mid s_q);\alpha_q) - \alpha_q \mathcal{H}_{\min} \right],
	% 	$}
	% \end{equation}
	%where ${\pi _q^ * \left( {{a_q}\left| {{s_q}} \right.} \right)}$ denotes the optimal policy corresponding to $\alpha_q$, and $\mathcal H_{\rm min}$ represents the minimum entropy constraint at each time slot.
	
	The basic structure of SAC builds on the policy iteration algorithm, which consists of policy evaluation and policy improvement phases. In the policy evaluation phase, the goal is to evaluate the action values for a given policy $\pi$ using the Bellman expectation function, expressed as $Q_\pi(s_q, a_q) = r(s_q, a_q) + \gamma \mathbb E_{s_{q+1} \sim p_s} \left[ v_\pi(s_{q+1}) \right]$. Different from  conventional DRL algorithms, by incorporating the entropy, the state-value function of SAC is defined as
	\begin{equation}
		{v_\pi }\left( {{s_q}} \right) = {\mathbb E_{{a_q} \sim \pi }}\left[ {{Q_\pi }\left( {{s_q},{a_q}} \right) - \alpha {{\log }_2}\left( {\pi \left( {{a_q}\left| {{s_q}} \right.} \right)} \right)} \right].
	\end{equation}
	Fig. 2 illustrates the SAC training process.
	\begin{figure}[t]
		\centering
		\includegraphics[width=0.45\textwidth]{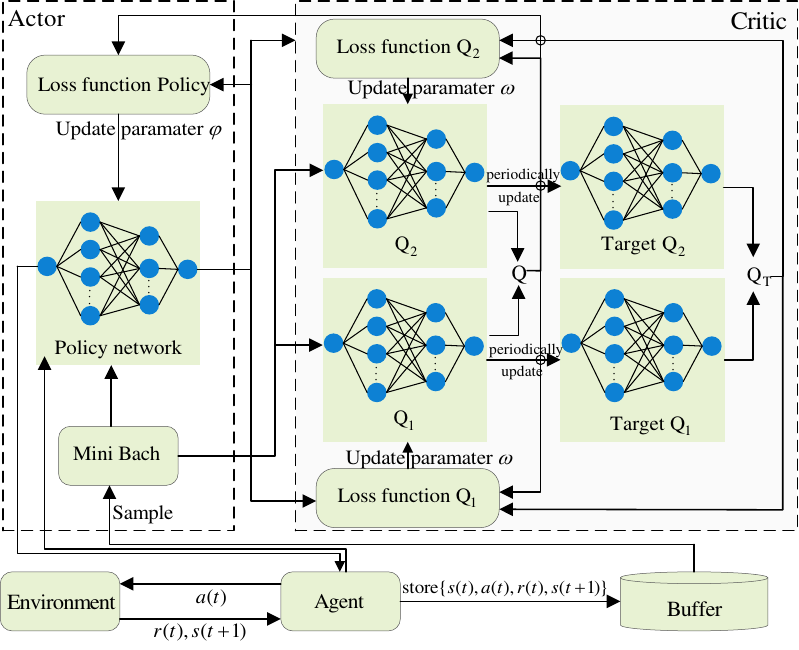}
		\caption{Training workflow of the SAC-based algorithm.}
		\label{fig:network}
	\end{figure}
	%\begin{algorithm}[t]
	%	\caption{SAC-based algorithm}
	%	\label{SAC}
	%	\begin{algorithmic}[1]
	%		\STATE{Initialize environment;}
	%		\STATE{
	%			Initialize ${\omega _i}(i = 1,2)$ for critic networks, $\varphi$ for actor network;}
	%		\STATE{
	%			Initialize entropy level $\mathcal H_{\min }$, replay buffer $D = \emptyset$, learning rate and temperature parameter respectively;}
	%		\FOR{each epoch}
	%		\FOR{each step}
	%%		\STATE{Calculate reward $r\left( s_q,a_q \right)$, $f_k^{\rm loc}$, $f_k^e$ by (21), (22), (23) respectively, obtain $p_k^{\rm off}$ by solving problem (26), and observe next state $s_{q+1}$;}
	%		\STATE{Store transition $\left\{ s_q, a_q, r(s_q,a_q), s_{q+1} \right\}$ in $\cal D$, and update UEs' location;}
	%		\FOR{each gradient step}
	%		\STATE{Randomly sample a mini-batch of transitions from $\cal D$;}
	%		\STATE{Update critic networks $\omega_i$ by minimizing loss function (31);}
	%		\STATE{Update the actor network $\varphi$ by minimizing loss function (32);}
	%		\STATE{Update temperature $\alpha$ by minimizing (29);}
	%		\STATE{Update target network parameter $\hat \omega_i$ periodically;}
	%		\ENDFOR
	%		\ENDFOR
	%		\RETURN{The optimal policy $\rm{\pi}_{\varphi}^{*}$}
	%	\end{algorithmic}
	%\end{algorithm}
	
	\section{simulation results}
	In this section, numerical results are presented to demonstrate the feasibility and effectiveness of the proposed algorithm. UEs are randomly positioned within a 5-meter radius circle centered at (30, 10, 0) meters, while the BS and RIS are located at (0, 0, 0) and (30, 0, 0) meters, respectively. The simulation parameters are detailed as follows: $p_{\max}=20 \text{ }\mathrm{dBm}$, $K=12$, $T=10\text{ }\mathrm{s}$, $Q = 5$, $B=12\text{ }\mathrm{MHz}$, $\sigma ^2=-110\text{ }\mathrm{dBm}$, $\alpha_1=\alpha_2=2$, $K_1=K_2=10\text{ }\mathrm{dB}$, $b=2\text{ }\mathrm{bit}$, $c_k^{\mathrm{loc}}=10^{-27}\text{ }\mathrm{J/cycle}$, ${D_k}=10\text{ }\mathrm{Mb}$, ${C_k}=600\text{ }\mathrm{cycles/bit}$, $f_{\mathrm{total}}^{\mathrm{e}}=10\text{ }\mathrm{GHz/s}$, $f_{\max}^{\mathrm{loc}}=0.6\text{ }\mathrm{GHz/s}$, $z=2$.
		\begin{figure}[t]
			\centering
			\includegraphics[width=0.95\columnwidth]{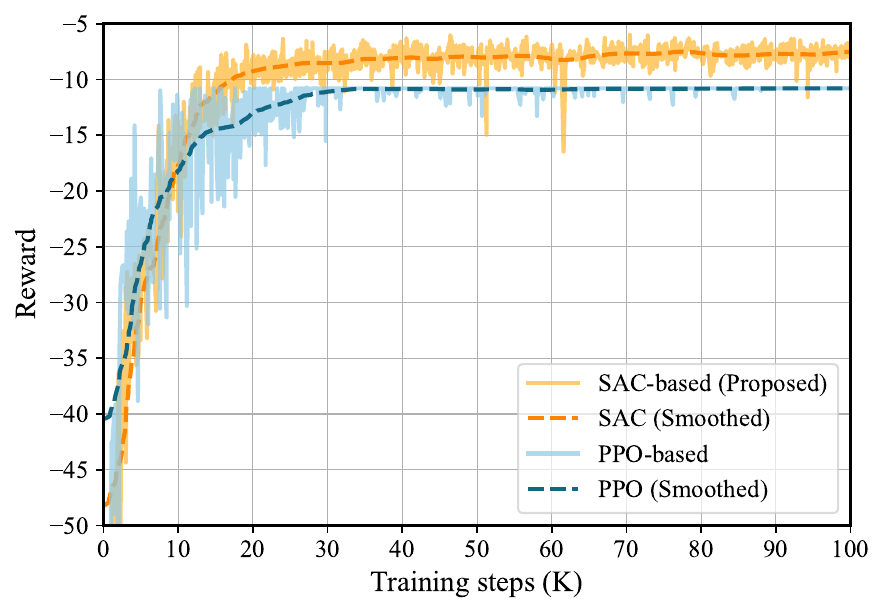}
			\caption{Convergence performance comparison for SAC and PPO, where $K=5$.}
			\label{fig:convergence}
		\end{figure}
	
			\begin{figure}[t]
				\centering
				\includegraphics[width=0.95\columnwidth]{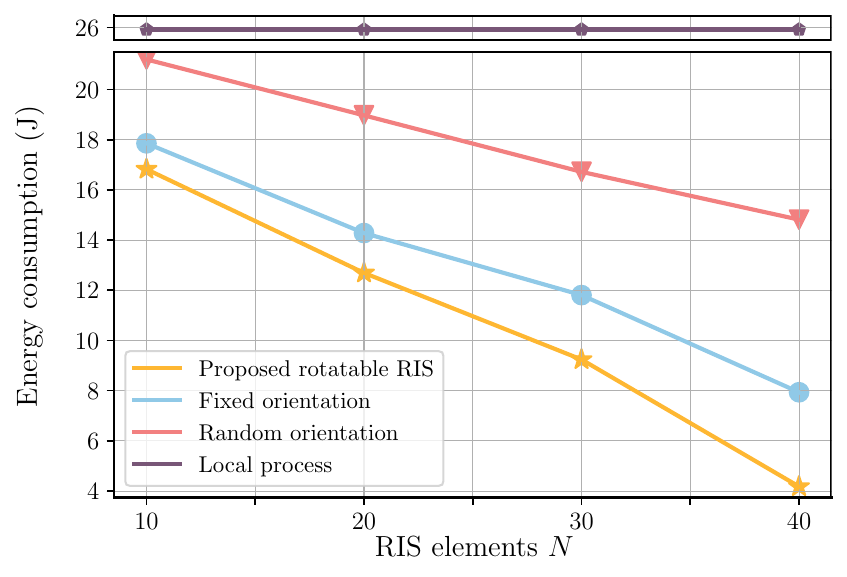}
			\caption{Performance VS the number of RIS elements, where $K=12$.}
			\label{fig:RISMUN}
			\end{figure}
				
			\begin{figure}[t]
				\centering
				\includegraphics[width=0.95\columnwidth]{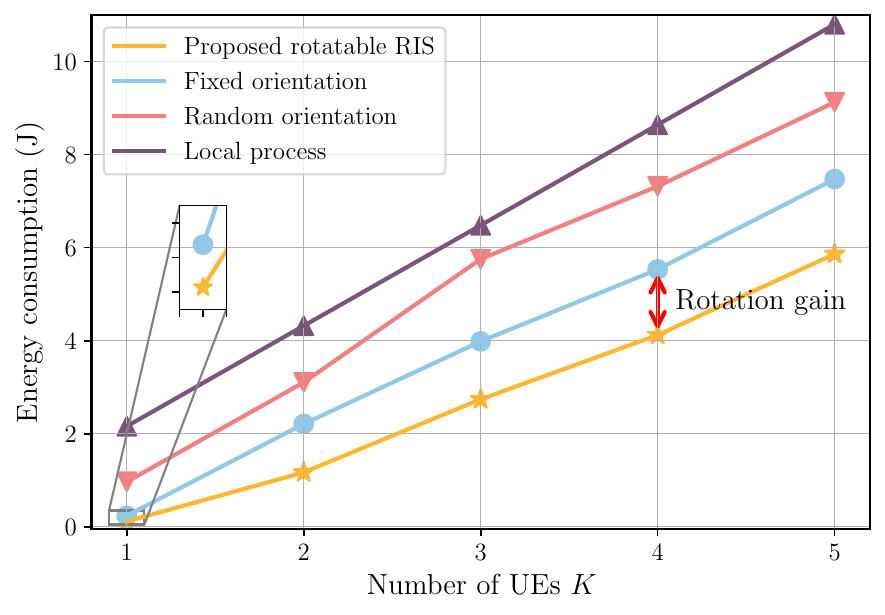}
			\caption{Performance VS UEs number, where $N=20$ and $B=$ 3 MHz.}
			\label{fig:USERNUM}
			\end{figure}

	\begin{figure*}[h]
		\centering
		\begin{subfigure}[t]{0.235\textwidth}
			\centerline{\includegraphics[width=\textwidth]{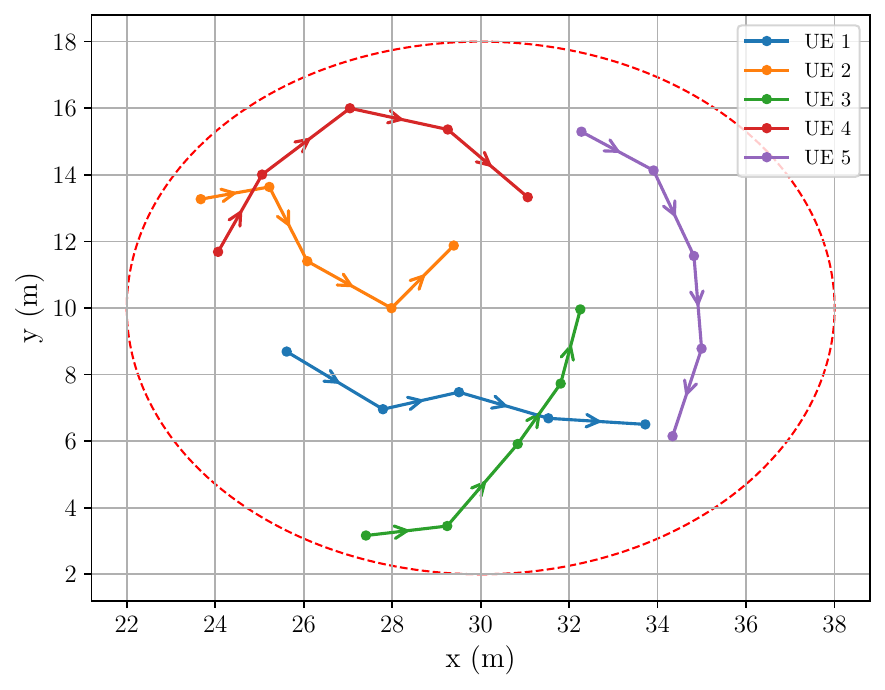}}
			\caption{UEs movement Trajectory.}
		\end{subfigure}
		\begin{subfigure}[t]{0.23\textwidth}
			\centerline{\includegraphics[width=\textwidth]{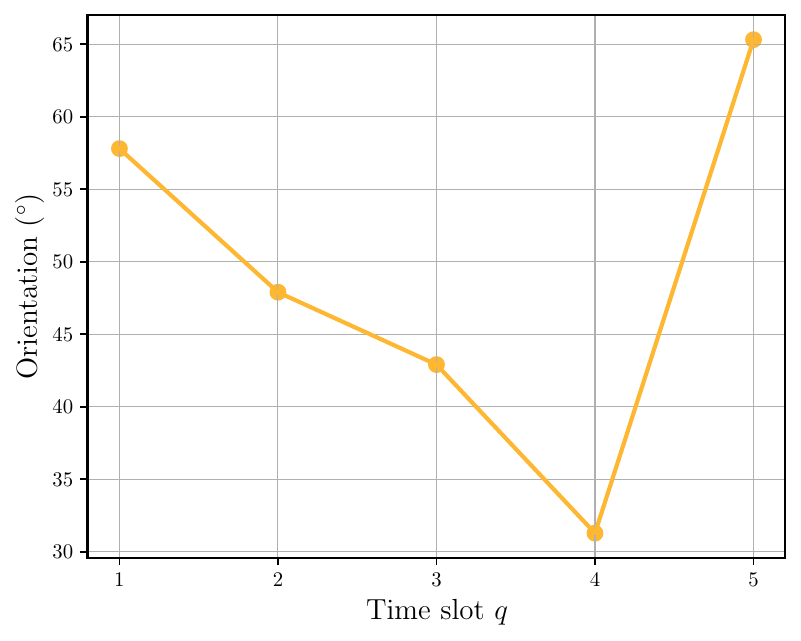}}
			\caption{The second time slot.}
		\end{subfigure}
		\begin{subfigure}[t]{0.24\textwidth}
			\centerline{\includegraphics[width=\textwidth]{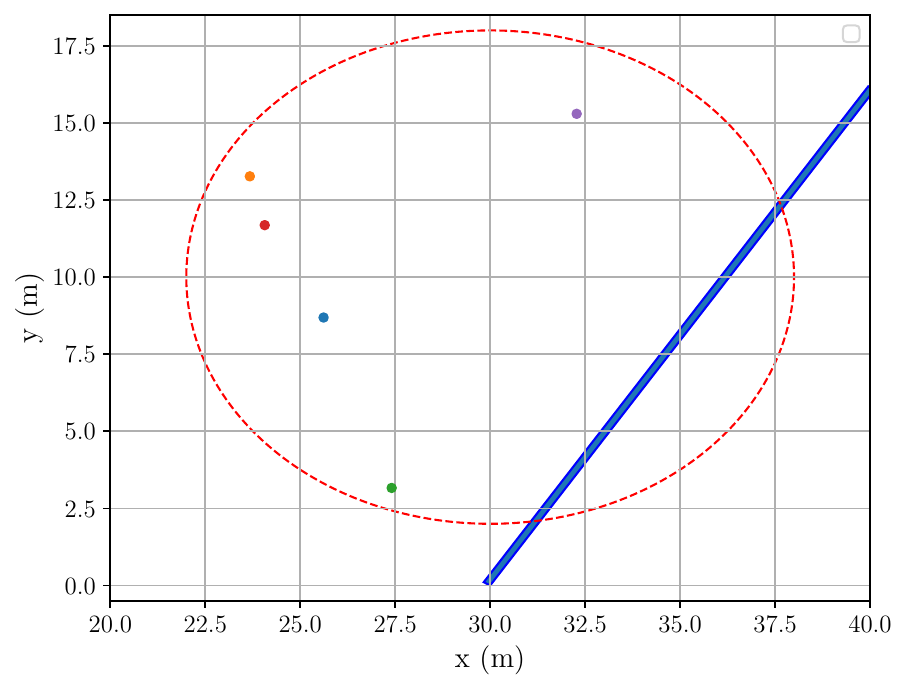}}
			\caption{The first time slot.}
		\end{subfigure}
		\begin{subfigure}[t]{0.24\textwidth}
			\centerline{\includegraphics[width=\textwidth]{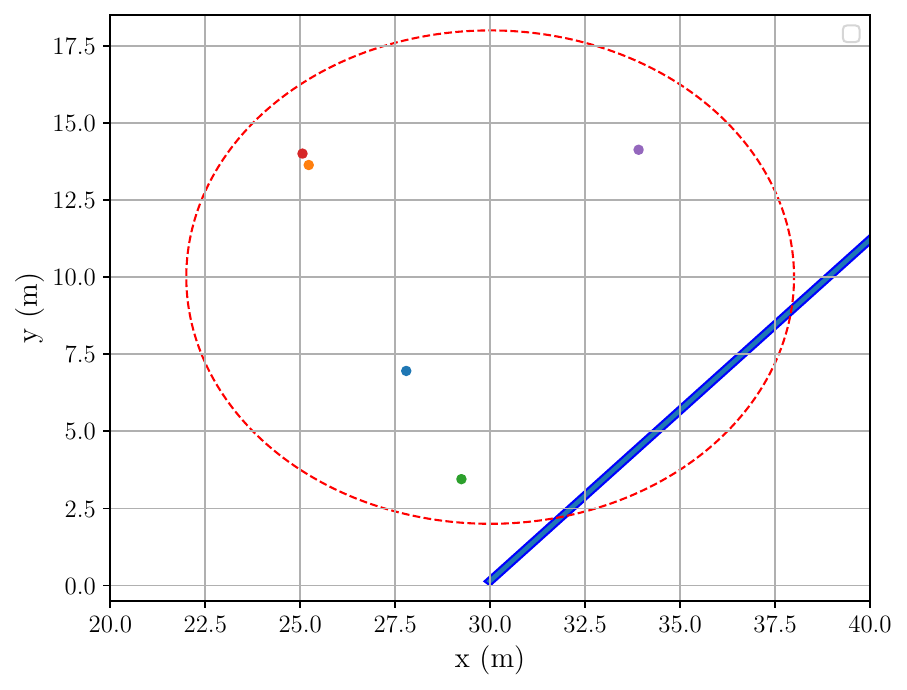}}
			\caption{The second time slot.}
		\end{subfigure}
		\begin{subfigure}[t]{0.24\textwidth}
			\centerline{\includegraphics[width=\textwidth]{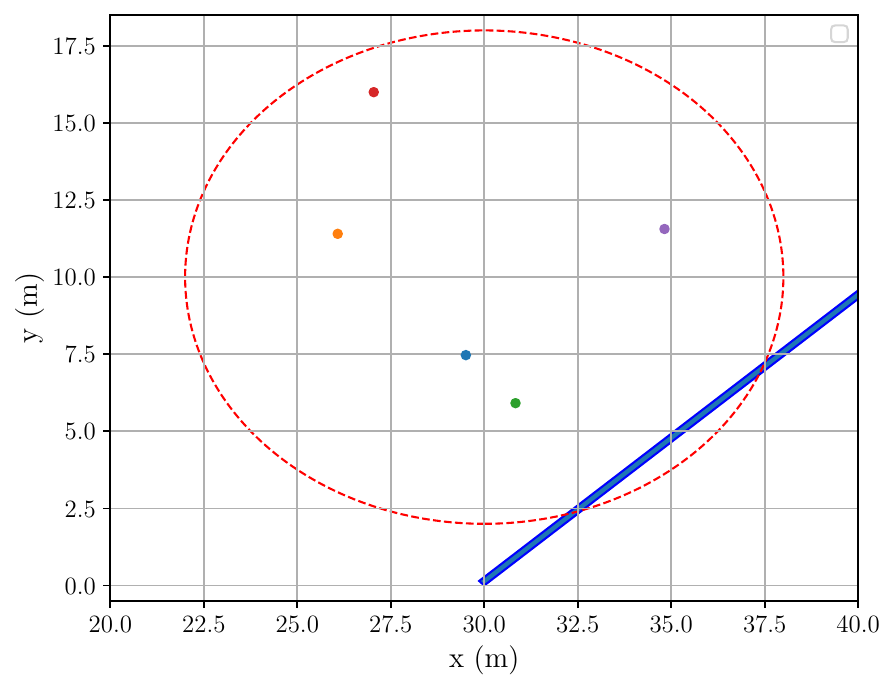}}
			\caption{The third time slot.}
		\end{subfigure}
		\begin{subfigure}[t]{0.24\textwidth}
			\centerline{\includegraphics[width=\textwidth]{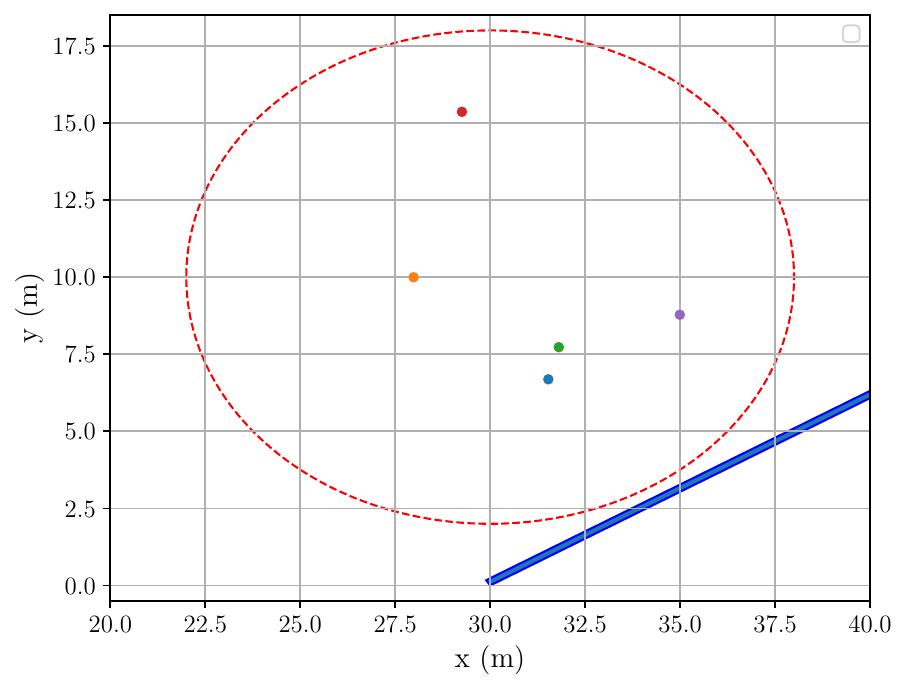}}
			\caption{The fourth time slot.}
		\end{subfigure}
		\begin{subfigure}[t]{0.24\textwidth}
			\centerline{\includegraphics[width=\textwidth]{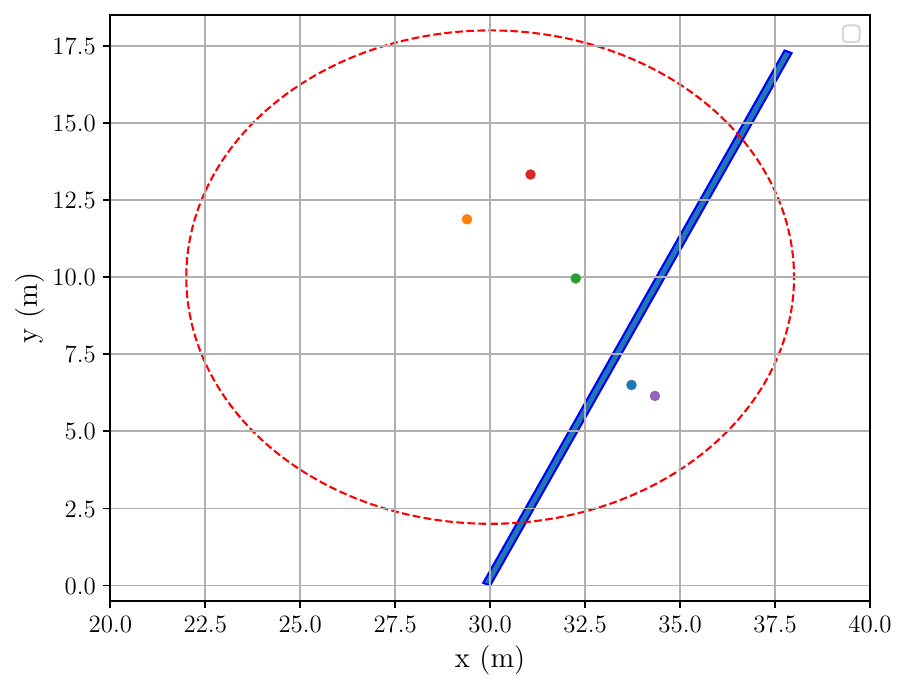}}
			\caption{The fifth time slot.}
		\end{subfigure}
		\begin{subfigure}[t]{0.24\textwidth}
			\centerline{\includegraphics[width=\textwidth]{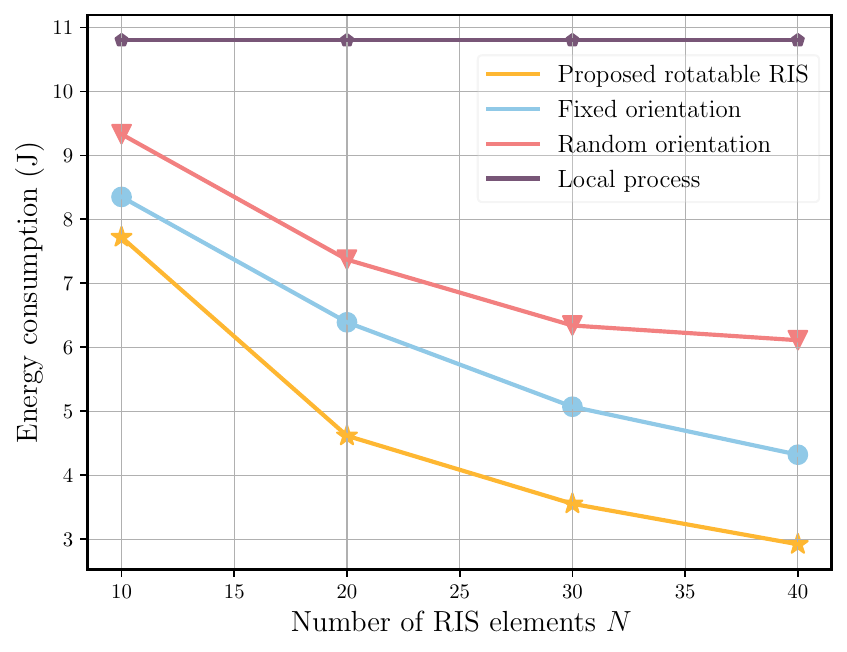}}
			\caption{The energy consumption VS RIS elements, with $K=5$.}
		\end{subfigure}
		\caption{The UEs movement trajectory, RIS orientation variation, and the energy consumption VS RIS elements.}
	\end{figure*}
	To validate the advantage of the proposed scheme, we compare it against four baseline schemes. $\romannumeral 1$) proximal policy optimization (PPO)-based algorithm; $\romannumeral 2$) Fixed orientation: the RIS orientation is fixed, maximizing the expected gain of the RIS; $\romannumeral 3$) Random orientation: the RIS orientation is randomly set; $\romannumeral 4$): UEs execute tasks locally without offloading, with $\alpha_k[q]=0$.
	
	In Fig. 3, we compare the convergence and performance of the SAC-based algorithm with the benchmark PPO-based algorithm. The proposed SAC-based algorithm demonstrates significant improvements in both convergence speed and overall performance. Within 20K training steps, the reward of the SAC-based algorithm rapidly increases and stabilizes, indicating a swift learning process and effective adaptation to the time-varying environment. In contrast, the PPO algorithm, although converging around 30K steps, reaches a worse reward value and stabilizes at 10.8, which corresponds to the total system power consumption when all UEs select local processing. Training data further indicates that the strategy provided by the PPO algorithm is local processing, mainly due to the set penalty and the insufficient optimization of the RIS phase shift,  which leads to weak channel gain and a preference for non-offloading. This comparison further highlights the stronger exploration capabilities of the SAC algorithm.

	Fig. 4 illustrates that UEs' energy consumption decreases as the number of RIS elements $N$ increases. This trend is expected since a larger $N$ facilitates an enhanced transmitting and receiving gain, contributing to a higher offloading ratio, and consequently lower energy consumption for UEs. Moreover, the energy consumption gap between the proposed rotatable RIS scheme and the fixed RIS scheme progressively widens, with the proposed scheme achieving a 47.3\% reduction when $N=40$. 
	
	To verify the effectiveness of the proposed scheme in scenarios with multiple moving UEs, we compare the performance of various schemes under different numbers of UEs. As shown in Fig. 5, the performance gap between the proposed scheme and other schemes widens as the number of UEs increases. Notably, the maximum improvement is observed when $K=1$, reaching $51.8$\%. This is because, with fewer UEs, the random movement poses  challenges for the fixed RIS to achieve the expected gain, whereas the rotatable RIS can dynamically adjust its orientation to better adapt to such scenarios.
	
	To further demonstrate the orientation adjustments of the proposed rotatable RIS scheme, we increase the UEs movement radius to $8$ m and set the number of UEs $K=5$. The UEs movement trajectory is presented in Fig. 6(a), while the corresponding orientation variations of the RIS are depicted in Fig. 6(b). To clearly depict the relationship between the UEs positions and RIS orientation, we plot the orientation of the RIS in each time slot relative to the UEs positions, as shown in Fig. 6(c)-(g). It can be observed that as the UEs move, the RIS continuously adjusts its orientation to optimize the RIS gain. In the first four time slots, all UEs are located within the reflection half-space of the RIS, and only two UEs fall outside in the last time slot. To investigate the impact of the number of RIS elements on performance gain, we compare the energy consumption under different numbers of RIS elements in Fig. 6(h). As the number of RIS elements $N$ increases, the total energy consumption steadily decreases. Moreover, the performance gap between our proposed rotatable RIS and the fixed RIS scheme becomes more pronounced as $N$ increases.
	
	\section{conclusion}
	This correspondence investigated the rotatable RIS-assisted MEC system with moving UEs, focusing on the joint optimization of the RIS orientation, discrete phase shift, computation resource allocation, transmitting power and task offloading strategies. We proposed an SAC-based algorithm that employs a stochastic strategy with entropy regulation, enabling near-optimal policy learning with minimal environmental information. Numerical results demonstrated that our SAC-based approach outperforms representative DRL algorithms in terms of convergence and overall performance. Additionally, lower energy consumption can be obtained by dynamically adjusting orientation according to the positions of BS and UEs.

	% 导入参考文献
	\bibliographystyle{IEEEtran}
	\bibliography{reference5.bib}  % references.bib 是你的 BibTeX 文件名
\end{document}